\documentclass[runningheads]{llncs}

\newcommand\tuple[1]{\langle #1 \rangle}
\newcommand{\GG}{{\cal G}}
\newcommand{\UU}{{\cal U}}

\newcommand{\pats}{\textsf{pats}}
\newcommand{\pat}{\textsf{pat}}

\newcommand{\etal}{\textit{et al.\ }}

\newcommand{\set}[1]{\left\{
      \begin{array}{l}#1\end{array}
      \right\}}
\newcommand{\sset}[2]{\left\{~#1  \left|
      \begin{array}{l}#2\end{array}
    \right.     \right\}}
\newcommand{\false}{\mathit{false}}
\newcommand{\true}{\mathit{true}}

\usepackage{cite}
\usepackage{tikz}
\usepackage{cite}
\usepackage{amsmath}
\usepackage{amssymb}
\newtheorem{observation}{Observation}

\usepackage{algorithm}
\usepackage{bwmatrix} %

\usepackage[noEnd=true,indLines=true]{algpseudocodex}
\usepackage{wrapfig}

\title{Breaking Symmetries from \\a Set-Covering Perspective}
\author{%
  Michael Codish\inst{1}\orcidID{0000-0003-0394-5854}
\and Mikol{\'{a}}\v{s} Janota\inst{2}\orcidID{0000-0003-3487-784X}} \institute{
  Department of Computer Science\\
  Ben-Gurion University of the Negev\\
  Beer-Sheva, Israel \\
  \email{\{mcodish\}@bgu.ac.il}
  \and
  Czech Technical University in Prague, Czechia
}
\authorrunning{M. Codish \and M. Janota}

\usepackage{color}
\definecolor{citeblue}{rgb}{0.1,0,.4}
\usepackage[pdftex%
,colorlinks=true%
,bookmarks=true%
,linkcolor=citeblue%
,citecolor=citeblue%
,plainpages=false]{hyperref}

\begin{document}
\maketitle

\begin{abstract}
  We formalize symmetry breaking as a set-covering problem. For the
  case of breaking symmetries on graphs, a permutation covers a graph
  if applying it to the graph yields a smaller graph in a given
  order. Canonical graphs are those that cannot be made smaller by any
  permutation. A complete symmetry break is then a set of permutations
  that covers all non-canonical graphs. A complete symmetry break with
  a minimal number of permutations can be obtained by solving an
  optimal set-covering problem.
  The challenge is in the sizes of the corresponding set-covering
  problems and in how these can be tamed.
  The set-covering perspective on symmetry breaking opens up a range
  of new opportunities deriving from decades of studies on both
  precise and approximate techniques for this problem.
  Application of our approach leads to optimal \textsc{LexLeader}
  symmetry breaks for graphs of order $n\leq 10$ as well as to
  partial symmetry breaks which improve on the state-of-the-art.
\end{abstract}

\section{Introduction}

Graph search problems are about finding simple graphs with desired
structural properties.
Such problems arise in many real-world applications and are 
fundamental in graph theory.
Solving graph search problems is typically hard due to the enormous
search space and the large number of symmetries in graph
representation.
For graph search problems, any graph obtained by permuting the
vertices of a solution (or a non-solution) is also a solution (or a
non-solution), which is isomorphic, or ``symmetric''.
When solving graph search problems, the presence of an enormous
number of symmetries typically causes redundant search effort by
revisiting symmetric objects.
To optimize the search we aim to restrict it to focus on one
``canonical'' graph from each isomorphism class.

The focus on symmetry has facilitated the solution of many open
instances of combinatorial search problems and graph search problems
in particular.  For example, the proof that the Ramsey numbers
$R(3,3,4)$ and $R(4,5)$ are equal to 30~\cite{Ramsey334} and to
25~\cite{GauthierB24} respectively, the solution for the Sudoku
minimum number of clues problem~\cite{sudoku}, and the enumeration of
all non-word representable graphs of order
twelve~\cite{Itzhakov_Codish_2020}.

One common approach to eliminate symmetries is to add \emph{symmetry
  breaking constraints} which are satisfied by at least one member of
each isomorphism class~\cite{CrawfordGLR96,Shlyakhter07,Walsh06}.
A symmetry breaking constraint is called \emph{complete} if it is
satisfied by exactly one member of each isomorphism class and
\emph{partial} otherwise.

In many cases, symmetry breaking constraints, complete or partial, are
expressed in terms of conjunctions of ``lex-constraints''. Each
constraint corresponds to one symmetry, $\sigma$, which is a
permutation on vertices, and restricts the search space to consider
assignments that are lexicographically smaller than their permuted
form obtained according to $\sigma$.
Similarly, a set of permutations is identified with the
conjunction of the lex-constraints for the elements of the set. A
complete symmetry break is a set that satisfies exactly the set
of canonical graphs.
Of course, if one considers the set of all permutations, then the
corresponding symmetry break is complete but too large to be of
practical use.

Codish~\etal\cite{Codish2019} introduce a partial symmetry break, which
is equivalent to considering the quadratic number of permutations
that swap a pair of vertices. Rintanen~\etal\cite{RintanenR24}
enhance this approach for directed graphs. 
Itzhakov and Codish~\cite{Itzhakov2016} observe that a complete
symmetry break can be defined in terms of a small number of
lex-constraints. They compute compact complete symmetry breaking
constraints for graphs with 10 or less vertices.
Similar approach is taken by Dan\v{c}o~et~al.\ in the context of finite models~\cite{danco2025}.
It is known that
that breaking symmetry by adding constraints to eliminate symmetric solutions is intractable in general~\cite{Babai1983,crawford-kr96}.
So, we do not expect to find a complete
symmetry break of polynomial size that identifies canonical graphs
which are lex-leaders.

In general, previous works focus on sets of permutations. Given a set
$\Pi$ of permutations, one typically asks questions of the form:
``Which symmetries are broken by $\Pi$'', ``Are all symmetries broken
by $\Pi$?'', ``Can we add a permutation to $\Pi$ and break a symmetry
not yet broken?'', or ``Can we remove a permutation from $\Pi$ and
still break the same symmetries?''.

In this paper we take a different view.  We focus on individual
permutations. We say that a permutation covers a graph if its
application on the graph yields a smaller graph.  Each permutation
``covers'' a set of graphs. A complete symmetry break is a set of
permutations, the elements of which cover all non-canonical graphs.
One essential question is: ``Which permutations are \emph{essential}
because they alone cover some graph?'' We call such permutations
``backbones''.

The set-cover problem is a classical problem in computer science and
one of the 21 NP-complete problems presented in the seminal paper by
Karp~\cite{Karp1972} in 1972. 
Given a set of elements, $\UU$, called the universe, and a collection
$S$ of subsets of $\UU$ whose union equals $\UU$, the set-cover
problem is to identify a smallest sub-collection of $S$ whose union
equals $\UU$.

By viewing symmetry breaking as a set-cover problem we make available
a wide range of techniques which have been studied for many decades
and applied to find set-covers, both exact and approximate. For
example,
\cite{roth1969,Caprara2000,Hoffman2001,Liu2020,Lei2020,Gupta2023}.
The main challenge derives from the fact that in the set-cover
perspective of symmetry breaking: (a) the universe consists of all
non-canonical graphs which is a set of the order $2^{n^2}$,  and (b)
the number of subsets considered corresponds to the non-identity
permutations which is a set of order $n!$.

In this paper we focus on the search for optimal complete symmetry
breaks for graphs. These are derived as solutions to minimal set-cover
problems. To tame the size of the corresponding set-cover problems we
focus on three classic optimizations applied to set-cover
problems~\cite{roth1969}. In our context we call these: (1) graph
dominance, (2) permutation dominance, and (3) identification of
permutation backbones.

Heule also addresses the problem of computing optimal symmetry breaks for
graphs~\cite{DBLP:conf/synasc/Heule16,Heule2019}.  Heule seeks an
answer in terms of the number of clauses in a CNF representation of
the corresponding symmetry breaking constraint. For up to $n=5$
vertices, Heule computes CNF size-optimal compact and complete
symmetry breaks.
We aim to find symmetry breaks that are optimal in their number of
lex-constraints. We show that this can be done for all of the cases in
which there exist complete and compact symmetry breaks based on
lex-constraints. Namely, for graphs of orders $n\leq 10$.

\section{Preliminaries and Notation}

Throughout this paper we consider simple graphs, i.e.\ undirected
graphs with no self-loops.
The adjacency matrix of a graph $G$ is an $n\times n$ Boolean
matrix. The element at row $i$ and column $j$ is $\true$ if and only
if $(i,j)$ is an edge. We denote by $vec(G)$ the sequence of length
$\binom{n}{2}$ which is the concatenation of the rows of the upper
triangle of $G$. In abuse of notation, we let $G$ denote a graph in
any of its representations.
The set of simple graphs on $n$ vertices is denoted~$\GG_n$.
An \emph{unknown graph} of order-$n$ is represented as an $n\times n$
adjacency matrix of Boolean variables which is symmetric and has the
values $\false$ (denoted by 0) on the diagonal.
We consider the following lexicographic ordering on graphs.
\begin{definition}[ordering graphs]\label{def:order}
  Let $G_1,G_2$ be known or unknown graphs with $n$ vertices and let
  $s_1=vec(G_1)$ and $s_2=vec(G_2)$ be the strings obtained by
  concatenating the rows of the upper triangular parts of their
  corresponding adjacency matrices.
  Then, $G_1 \leq G_2$ if and only if $s_1\leq_{lex} s_2$. 
\end{definition}
When $G_1$ and $G_2$ are unknown graphs, then the lexicographic
ordering, $G_1\leq G_2$, can be viewed as specifying a
\emph{lexicographic order constraint} over the variables in $vec(G_1)$
and $vec(G_2)$.

The group of permutations on $\{1 \ldots n\}$ is denoted $S_n$.
We represent a permutation $\pi \in S_n$ as a sequence of length $n$
where the i$^{th}$ element indicates the value of $\pi(i)$.  For
example: the permutation $[2,3,1] \in S_3$ maps as follows:
$\{1 \mapsto 2, 2 \mapsto 3, 3 \mapsto 1\}$.
Permutations act on graphs and on unknown graphs in the natural
way. For a graph $G\in\GG_n$ and also for an unknown graph $G$,
viewing $G$ as an adjacency matrix, given a permutation $\pi\in S_n$,
then $\pi(G)$ is the adjacency matrix obtained by mapping each element
at position $(i,j)$ to position $(\pi(i),\pi(j))$ (for $1\leq i,j\leq
n$). Alternatively, $\pi(G)$ is the adjacency matrix obtained by
permuting both rows and columns of $G$ using $\pi$.
Two graphs $G,H\in\GG_n$ are \emph{isomorphic} if there exists a
permutation $\pi \in S_n$ such that $G=\pi(H)$.
\begin{example}\label{example:lex-constraints}

  The following depicts an unknown, order-4, graph $G$, its
  permutation $\pi(G)$, for $\pi = [1,2,4,3]$, and their vector
  representations.  The lex-constraint $G \leq \pi(G)$ can be
  simplified as described by Frisch~\etal\cite{Frisch03} to:
  $\tuple{x_2,x_4} \leq_{lex} \tuple{x_3,x_5}$.\medskip

  \noindent
  {\small\begin{tabular}{lll}
 {$\mathbf{G=}\left[\begin{matrix}
    0 & x_1 & x_2 & x_3 \\
    x_1 & 0 & x_4 & x_5 \\
    x_2 & x_4 & 0 & x_6 \\
    x_3 & x_5 & x_6 & 0 
\end{matrix}\right]$}
&
\quad  {$\mathbf{\pi(G)=}\left[\begin{matrix}
    0 & x_1 & x_3 & x_2 \\
    x_1 & 0 & x_5 & x_4 \\
    x_3 & x_5 & 0 & x_6 \\
    x_2 & x_4 & x_6 & 0 
\end{matrix}\right]$}
&
\quad 
$\begin{array}{rll}
        vec(G)&=&\tuple{x_1,x_2,x_3,x_4,x_5,x_6}\\  
        vec(\pi(G))&=&\tuple{x_1,x_3,x_2,x_5,x_4,x_6}
\end{array}$
\end{tabular}
         }
\end{example}

A \emph{graph search problem} is a predicate, $\varphi(G)$, on an unknown
graph $G$, which is invariant under the names of vertices. In other
words, it is invariant under isomorphism. A solution to $\varphi(G)$
is a satisfying assignment for the variables of~$G$.
Graph search problems include existence problems, where the goal is to
determine whether a simple graph with certain graph properties exists,
enumeration problems, which are about finding all solutions (modulo
graph isomorphism), and extremal problems, where we seek the
smallest/largest solution with respect to some target (such as the
number of edges or vertices in a solution).
Solving graph search problems is typically hard due to the enormous
search space and the large number of symmetries.

A symmetry break for graph search problems is a predicate, $\psi(G)$,
on a graph $G$, which is satisfied by at least one graph in each
isomorphism class of graphs. If $\psi$ is satisfied by exactly one
graph in each isomorphism class then we say that $\psi$ is a complete
symmetry break. Otherwise it is partial.
A classic complete symmetry break for graphs is the \textsc{LexLeader}
constraint~\cite{Read1978} defined as follows:
\begin{definition}[\textsc{LexLeader}]
  Let $G$ be an unknown order-$n$ graph, Then,
  \begin{equation}\label{lexleader}
    \textsc{LexLeader}(n) = 
    \bigwedge\sset{G \leq\pi(G)}{\pi\in S_n}
  \end{equation}
  
\end{definition}

The \textsc{LexLeader} constraint is impractical as it is composed of
a super-exponential number of constraints, one for each permutation of
the vertices.
A symmetry break which is equivalent to the \textsc{LexLeader}
constraint is called a \textsc{LexLeader} symmetry break.
In~\cite{Itzhakov2016}, the authors present a methodology to compute 
\textsc{LexLeader} symmetry breaks which are much smaller in size.

\begin{definition}[canonizing set of permutations]\label{def:canonizing}
  Let $\Pi \subseteq S_n$ be a set of permutations such that
  \[\forall G.~ \textsc{LexLeader}(n)(G) \Leftrightarrow
    \bigwedge_{\pi \in \Pi} G  \leq \pi(G)
  \]
  In this case we say that $\Pi$ is a canonizing set of permutations.  
\end{definition}

In a nutshell, the algorithm presented in~\cite{Itzhakov2016} computes
a canonizing set $\Pi$ initialized to the empty set by incrementally
performing $\Pi\gets\{\pi'\} \cup \Pi$ as long as there exists a
permutation $\pi'$ and a graph $G$ such that
\[\bigwedge_{\pi\in\Pi} G\leq\pi(G) ~\wedge~ \pi'(G)< G\]
\begin{wraptable}{r}{4.8cm}
  \begin{tabular}{|c|c|c|c|c|c|c|c|c|}
  \hline
  \textsf{order} & 3 & 4 & 5 & 6  & 7  & 8   & 9   & 10   \\
  \hline
  \textsf{size}  & 2 & 3 & 7 & 13 & 37 & 135 & 842 & 7853 \\
  \hline
  \end{tabular}
  \vspace{-8mm}
\end{wraptable}
There are additional subtleties related to the removal of ``redundant''
permutations from the resulting set $\Pi$ (see~\cite{Itzhakov2016}).
The canonizing sets obtained are surprisingly
small. Their sizes are detailed on the right.

\section{The Set-Covering Perspective}\label{section:setCover}

 The set-cover perspective on symmetry breaking is based on the notion
 that a permutation covers the set of graphs that it makes smaller.

\begin{definition}[cover]\label{def:cover}\\
  Let $\pi\in S_n$. Then, $\pi$ covers an order-$n$ graph $G$ if
  $\pi(G)<G$. We denote the set
  $cover(\pi) = \sset{G}{\pi(G)<G}$. For a set of permutations
  $\Pi\subseteq S_n$, we denote the set
  $cover(\Pi) = \sset{G}{\pi\in\Pi, \pi(G)<G}$.
\end{definition}

Let $\Pi$ be a canonizing set of permutations. It follows from
Def.~\ref{def:canonizing} that a graph $G$ is canonical if and
only if the constraints $G \leq \pi(G)$ hold for all
$\pi\in\Pi$.  
The contra-positive states that $G$ is non-canonical, if and only if
there exists $\pi\in\Pi$ such that $\pi(G)<G$. Hence, we can
view a canonizing set $\Pi$ as a set such that for for every
non-canonical graph $G$, $\Pi$ contains a permutation $\pi$ such that
$\pi$ covers $G$. We can state this as follows:

\begin{observation}\label{obs:contrapositive}
  A set of permutations is canonizing if and only if it covers all
  non-canonical graphs.
\end{observation}

While Observation~\ref{obs:contrapositive} might appear trivial, the
nature of the statement leads to a new and alternative view on symmetry
breaking in terms of set-covering.
The goal for complete symmetry breaking is to cover all non-canonical
graphs by a set of permutations. Of course, the set of all
permutations cover all non-canonical graphs. However we can seek a
smaller set of permutations and in particular, a set of the smallest
size.

\begin{definition}[optimal lex-constraint symmetry break]
  An optimal lex-constraint symmetry break is a canonizing set of
  permutations of minimal cardinality. The optimal lex-constraint
  symmetry break problem is that of finding an optimal lex-constraint
  symmetry break.
\end{definition}

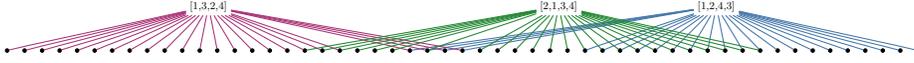
\begin{figure}[tb]
  \centering

\definecolor{ptpurple}{RGB}{ 170,51,119}
\definecolor{ptgreen}{RGB}{34,136,51}
\definecolor{ptblue}{RGB}{68,119,170}
\begin{tikzpicture}[scale=.3,>=latex,line join=bevel,%
    ,every node/.style={scale=.15}]
\node (g2) at (727.8bp,1.8bp) [draw,circle, fill] {};
  \node (g3) at (749.8bp,1.8bp) [draw,circle, fill] {};
  \node (g8) at (1035.8bp,1.8bp) [draw,circle, fill] {};
  \node (g9) at (551.8bp,1.8bp) [draw,circle, fill] {};
  \node (g10) at (1057.8bp,1.8bp) [draw,circle, fill] {};
  \node (g11) at (1079.8bp,1.8bp) [draw,circle, fill] {};
  \node (g14) at (771.8bp,1.8bp) [draw,circle, fill] {};
  \node (g15) at (793.8bp,1.8bp) [draw,circle, fill] {};
  \node (g18) at (815.8bp,1.8bp) [draw,circle, fill] {};
  \node (g19) at (529.8bp,1.8bp) [draw,circle, fill] {};
  \node (g26) at (1101.8bp,1.8bp) [draw,circle, fill] {};
  \node (g27) at (573.8bp,1.8bp) [draw,circle, fill] {};
  \node (g34) at (837.8bp,1.8bp) [draw,circle, fill] {};
  \node (g35) at (859.8bp,1.8bp) [draw,circle, fill] {};
  \node (g40) at (1123.8bp,1.8bp) [draw,circle, fill] {};
  \node (g41) at (507.8bp,1.8bp) [draw,circle, fill] {};
  \node (g42) at (1145.8bp,1.8bp) [draw,circle, fill] {};
  \node (g43) at (969.8bp,1.8bp) [draw,circle, fill] {};
  \node (g46) at (881.8bp,1.8bp) [draw,circle, fill] {};
  \node (g47) at (903.8bp,1.8bp) [draw,circle, fill] {};
  \node (g50) at (925.8bp,1.8bp) [draw,circle, fill] {};
  \node (g51) at (947.8bp,1.8bp) [draw,circle, fill] {};
  \node (g58) at (991.8bp,1.8bp) [draw,circle, fill] {};
  \node (g59) at (1013.8bp,1.8bp) [draw,circle, fill] {};
  \node (g1) at (243.8bp,1.8bp) [draw,circle, fill] {};
  \node (g5) at (375.8bp,1.8bp) [draw,circle, fill] {};
  \node (g13) at (265.8bp,1.8bp) [draw,circle, fill] {};
  \node (g16) at (287.8bp,1.8bp) [draw,circle, fill] {};
  \node (g17) at (309.8bp,1.8bp) [draw,circle, fill] {};
  \node (g20) at (331.8bp,1.8bp) [draw,circle, fill] {};
  \node (g21) at (353.8bp,1.8bp) [draw,circle, fill] {};
  \node (g23) at (397.8bp,1.8bp) [draw,circle, fill] {};
  \node (g24) at (1.8bp,1.8bp) [draw,circle, fill] {};
  \node (g25) at (23.8bp,1.8bp) [draw,circle, fill] {};
  \node (g28) at (45.8bp,1.8bp) [draw,circle, fill] {};
  \node (g29) at (67.8bp,1.8bp) [draw,circle, fill] {};
  \node (g31) at (89.8bp,1.8bp) [draw,circle, fill] {};
  \node (g33) at (111.8bp,1.8bp) [draw,circle, fill] {};
  \node (g37) at (419.8bp,1.8bp) [draw,circle, fill] {};
  \node (g45) at (133.8bp,1.8bp) [draw,circle, fill] {};
  \node (g49) at (155.8bp,1.8bp) [draw,circle, fill] {};
  \node (g53) at (177.8bp,1.8bp) [draw,circle, fill] {};
  \node (g57) at (199.8bp,1.8bp) [draw,circle, fill] {};
  \node (g61) at (221.8bp,1.8bp) [draw,circle, fill] {};
  \node (g4) at (683.8bp,1.8bp) [draw,circle, fill] {};
  \node (g6) at (705.8bp,1.8bp) [draw,circle, fill] {};
  \node (g7) at (441.8bp,1.8bp) [draw,circle, fill] {};
  \node (g22) at (463.8bp,1.8bp) [draw,circle, fill] {};
  \node (g36) at (485.8bp,1.8bp) [draw,circle, fill] {};
  \node (g38) at (595.8bp,1.8bp) [draw,circle, fill] {};
  \node (g39) at (617.8bp,1.8bp) [draw,circle, fill] {};
  \node (g54) at (639.8bp,1.8bp) [draw,circle, fill] {};
  \node (g55) at (661.8bp,1.8bp) [draw,circle, fill] {};
  \node (p1) at (892.8bp,57.6bp) [scale=3,text opacity=1,draw=none] {[1,2,4,3]};
  \node (p2) at (254.8bp,57.6bp) [scale=3,text opacity=1,draw=none] {[1,3,2,4]};
  \node (p6) at (694.8bp,57.6bp) [scale=3,text opacity=1,draw=none] {[2,1,3,4]};
  \draw [ptblue,solid] (p1) ..controls (809.16bp,29.327bp) and (739.74bp,6.6935bp)  .. (g2);
  \draw [ptblue,solid] (p1) ..controls (815.72bp,27.602bp) and (761.24bp,7.1054bp)  .. (g3);
  \draw [ptblue,solid] (p1) ..controls (969.88bp,27.602bp) and (1024.4bp,7.1054bp)  .. (g8);
  \draw [ptblue,solid] (p1) ..controls (768.83bp,37.041bp) and (571.48bp,5.9053bp)  .. (g9);
  \draw [ptblue,solid] (p1) ..controls (976.44bp,29.327bp) and (1045.9bp,6.6935bp)  .. (g10);
  \draw [ptblue,solid] (p1) ..controls (982.47bp,30.802bp) and (1066.8bp,6.527bp)  .. (g11);
  \draw [ptblue,solid] (p1) ..controls (822.39bp,25.293bp) and (780.56bp,6.6943bp)  .. (g14);
  \draw [ptblue,solid] (p1) ..controls (835.19bp,25.293bp) and (800.97bp,6.6943bp)  .. (g15);
  \draw [ptblue,solid] (p1) ..controls (848.7bp,25.786bp) and (823.2bp,7.9718bp)  .. (g18);
  \draw [ptblue,solid] (p1) ..controls (765.25bp,37.696bp) and (551.16bp,5.9659bp)  .. (g19);
  \draw [ptblue,solid] (p1) ..controls (988.29bp,32.021bp) and (1088.3bp,6.2815bp)  .. (g26);
  \draw [ptblue,solid] (p1) ..controls (773.64bp,36.503bp) and (594.82bp,6.345bp)  .. (g27);
  \draw [ptblue,solid] (p1) ..controls (861.3bp,25.786bp) and (843.09bp,7.9718bp)  .. (g34);
  \draw [ptblue,solid] (p1) ..controls (873.9bp,25.786bp) and (862.97bp,7.9718bp)  .. (g35);
  \draw [ptblue,solid] (p1) ..controls (993.11bp,33.237bp) and (1108.2bp,6.4296bp)  .. (g40);
  \draw [ptblue,solid] (p1) ..controls (761.07bp,38.192bp) and (530.78bp,6.0116bp)  .. (g41);
  \draw [ptblue,solid] (p1) ..controls (997.94bp,34.241bp) and (1128.2bp,6.5522bp)  .. (g42);
  \draw [ptblue,solid] (p1) ..controls (936.9bp,25.786bp) and (962.4bp,7.9718bp)  .. (g43);
  \draw [ptblue,solid] (p1) ..controls (886.5bp,25.786bp) and (882.86bp,7.9718bp)  .. (g46);
  \draw [ptblue,solid] (p1) ..controls (899.1bp,25.786bp) and (902.74bp,7.9718bp)  .. (g47);
  \draw [ptblue,solid] (p1) ..controls (911.7bp,25.786bp) and (922.63bp,7.9718bp)  .. (g50);
  \draw [ptblue,solid] (p1) ..controls (924.3bp,25.786bp) and (942.51bp,7.9718bp)  .. (g51);
  \draw [ptblue,solid] (p1) ..controls (950.41bp,25.293bp) and (984.63bp,6.6943bp)  .. (g58);
  \draw [ptblue,solid] (p1) ..controls (963.21bp,25.293bp) and (1005.0bp,6.6943bp)  .. (g59);
  \draw [ptpurple,solid] (p2) ..controls (369.29bp,35.861bp) and (532.61bp,6.276bp)  .. (g9);
  \draw [ptpurple,solid] (p2) ..controls (364.87bp,35.066bp) and (512.47bp,6.1907bp)  .. (g19);
  \draw [ptpurple,solid] (p2) ..controls (373.96bp,36.503bp) and (552.78bp,6.345bp)  .. (g27);
  \draw [ptpurple,solid] (p2) ..controls (359.94bp,34.241bp) and (490.16bp,6.5522bp)  .. (g41);
  \draw [ptpurple,solid] (p2) ..controls (248.5bp,25.786bp) and (244.86bp,7.9718bp)  .. (g1);
  \draw [ptpurple,solid] (p2) ..controls (325.21bp,25.293bp) and (367.04bp,6.6943bp)  .. (g5);
  \draw [ptpurple,solid] (p2) ..controls (261.1bp,25.786bp) and (264.74bp,7.9718bp)  .. (g13);
  \draw [ptpurple,solid] (p2) ..controls (273.7bp,25.786bp) and (284.63bp,7.9718bp)  .. (g16);
  \draw [ptpurple,solid] (p2) ..controls (286.3bp,25.786bp) and (304.51bp,7.9718bp)  .. (g17);
  \draw [ptpurple,solid] (p2) ..controls (298.9bp,25.786bp) and (324.4bp,7.9718bp)  .. (g20);
  \draw [ptpurple,solid] (p2) ..controls (312.41bp,25.293bp) and (346.63bp,6.6943bp)  .. (g21);
  \draw [ptpurple,solid] (p2) ..controls (331.88bp,27.602bp) and (386.36bp,7.1054bp)  .. (g23);
  \draw [ptpurple,solid] (p2) ..controls (149.66bp,34.241bp) and (19.445bp,6.5522bp)  .. (g24);
  \draw [ptpurple,solid] (p2) ..controls (154.49bp,33.237bp) and (39.384bp,6.4296bp)  .. (g25);
  \draw [ptpurple,solid] (p2) ..controls (159.31bp,32.021bp) and (59.325bp,6.2815bp)  .. (g28);
  \draw [ptpurple,solid] (p2) ..controls (165.13bp,30.802bp) and (80.754bp,6.527bp)  .. (g29);
  \draw [ptpurple,solid] (p2) ..controls (171.16bp,29.327bp) and (101.74bp,6.6935bp)  .. (g31);
  \draw [ptpurple,solid] (p2) ..controls (177.72bp,27.602bp) and (123.24bp,7.1054bp)  .. (g33);
  \draw [ptpurple,solid] (p2) ..controls (338.44bp,29.327bp) and (407.86bp,6.6935bp)  .. (g37);
  \draw [ptpurple,solid] (p2) ..controls (184.39bp,25.293bp) and (142.56bp,6.6943bp)  .. (g45);
  \draw [ptpurple,solid] (p2) ..controls (197.19bp,25.293bp) and (162.97bp,6.6943bp)  .. (g49);
  \draw [ptpurple,solid] (p2) ..controls (210.7bp,25.786bp) and (185.2bp,7.9718bp)  .. (g53);
  \draw [ptpurple,solid] (p2) ..controls (223.3bp,25.786bp) and (205.09bp,7.9718bp)  .. (g57);
  \draw [ptpurple,solid] (p2) ..controls (235.9bp,25.786bp) and (224.97bp,7.9718bp)  .. (g61);
  \draw [ptgreen,solid] (p6) ..controls (713.7bp,25.786bp) and (724.63bp,7.9718bp)  .. (g2);
  \draw [ptgreen,solid] (p6) ..controls (726.3bp,25.786bp) and (744.51bp,7.9718bp)  .. (g3);
  \draw [ptgreen,solid] (p6) ..controls (738.9bp,25.786bp) and (764.4bp,7.9718bp)  .. (g14);
  \draw [ptgreen,solid] (p6) ..controls (752.41bp,25.293bp) and (786.63bp,6.6943bp)  .. (g15);
  \draw [ptgreen,solid] (p6) ..controls (765.21bp,25.293bp) and (807.04bp,6.6943bp)  .. (g18);
  \draw [ptgreen,solid] (p6) ..controls (611.16bp,29.327bp) and (541.74bp,6.6935bp)  .. (g19);
  \draw [ptgreen,solid] (p6) ..controls (771.88bp,27.602bp) and (826.36bp,7.1054bp)  .. (g34);
  \draw [ptgreen,solid] (p6) ..controls (778.44bp,29.327bp) and (847.86bp,6.6935bp)  .. (g35);
  \draw [ptgreen,solid] (p6) ..controls (784.47bp,30.802bp) and (868.85bp,6.527bp)  .. (g46);
  \draw [ptgreen,solid] (p6) ..controls (790.29bp,32.021bp) and (890.28bp,6.2815bp)  .. (g47);
  \draw [ptgreen,solid] (p6) ..controls (795.11bp,33.237bp) and (910.22bp,6.4296bp)  .. (g50);
  \draw [ptgreen,solid] (p6) ..controls (799.94bp,34.241bp) and (930.16bp,6.5522bp)  .. (g51);
  \draw [ptgreen,solid] (p6) ..controls (575.64bp,36.503bp) and (396.82bp,6.345bp)  .. (g5);
  \draw [ptgreen,solid] (p6) ..controls (580.31bp,35.861bp) and (416.99bp,6.276bp)  .. (g23);
  \draw [ptgreen,solid] (p6) ..controls (584.73bp,35.066bp) and (437.13bp,6.1907bp)  .. (g37);
  \draw [ptgreen,solid] (p6) ..controls (688.5bp,25.786bp) and (684.86bp,7.9718bp)  .. (g4);
  \draw [ptgreen,solid] (p6) ..controls (701.1bp,25.786bp) and (704.74bp,7.9718bp)  .. (g6);
  \draw [ptgreen,solid] (p6) ..controls (589.66bp,34.241bp) and (459.44bp,6.5522bp)  .. (g7);
  \draw [ptgreen,solid] (p6) ..controls (594.49bp,33.237bp) and (479.38bp,6.4296bp)  .. (g22);
  \draw [ptgreen,solid] (p6) ..controls (599.31bp,32.021bp) and (499.32bp,6.2815bp)  .. (g36);
  \draw [ptgreen,solid] (p6) ..controls (637.19bp,25.293bp) and (602.97bp,6.6943bp)  .. (g38);
  \draw [ptgreen,solid] (p6) ..controls (650.7bp,25.786bp) and (625.2bp,7.9718bp)  .. (g39);
  \draw [ptgreen,solid] (p6) ..controls (663.3bp,25.786bp) and (645.09bp,7.9718bp)  .. (g54);
  \draw [ptgreen,solid] (p6) ..controls (675.9bp,25.786bp) and (664.97bp,7.9718bp)  .. (g55);
\end{tikzpicture}
	\caption{Optimal set-cover for $n=4$}%
	\label{fig:optcov4}
\end{figure}

\begin{figure}[t]
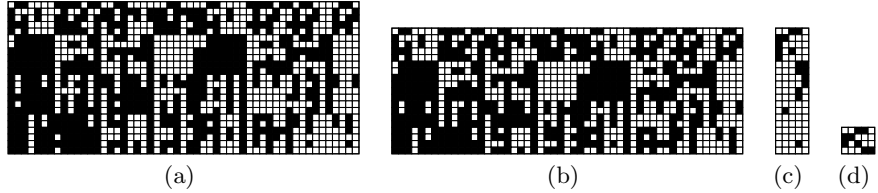

\begin{tabular}{cccc}
\plot{data_4_1.dat}{0.22}{0.4}{0.4}{0} & %
\plot{data_4_2.dat}{0.22}{0.4}{0.4}{0} & %
\plot{data_4_3.dat}{0.22}{0.4}{0.4}{0} &  %
\plot{data_4_4.dat}{0.22}{0.4}{0.4}{0} \\
  (a) & (b) & (c) & (d) \\                                                                                 
\end{tabular}
\caption{Optimal Symmetry break for $n=4$ as a set-cover problem}\label{data4_1}
\end{figure}

\begin{example}\label{ex:n4}
  Consider the case for graphs of order-4.  Each graph $G$ is
  represented as the integer value corresponding to the six digit
  binary sequence $vec(G)$ (viewed lsb first). The 11 canonical graphs
  $\{0,12,30,32,44,48,52,56,60,62,63\}$ are not covered by any of the
  permutations.  We detail below the sets of graphs covered by three of
  the permutations. Fig.~\ref{fig:optcov4} illustrates that these
  three permutations cover all of the non-canonical graphs of
  order-4. This is an optimal cover.  The dots on the bottom represent
  an enumeration of the non-canonical graphs.

{\scriptsize
\begin{verbatim}
[1,2,4,3]:{2,3,8,9,10,11,14,15,18,19,26,27,34,35,40,41,42,43,46,47,50,51,58,59}
[1,3,2,4]:{1,5,9,13,16,17,19,20,21,23,24,25,27,28,29,31,33,37,41,45,49,53,57,61}
[2,1,3,4]:{2,3,4,5,6,7,14,15,18,19,22,23,34,35,36,37,38,39,46,47,50,51,54,55}
\end{verbatim}}
  
The matrix depicted as Fig.~\ref{data4_1}(a) depicts the set-cover
problem corresponding to the cover sets for all of the permutations
(excluding the row for the identity permutation). The rows correspond
to the permutations, the columns correspond to the graphs, a cell is
colored black if the corresponding permutation covers the
corresponding graph, and white otherwise.
\end{example}

Three optimizations apply to simplify a set-cover problem. We
present these, adapted to the context of symmetry breaking, from the
presentation in~\cite{roth1969}.

\smallskip\noindent{\bf Optimization 1 (permutation dominance):}
If permutation $\pi_1$ covers a subset of the graphs covered by
permutation $\pi_2$, then we say that $\pi_2$ dominates $\pi_1$.
In this case we discard the row corresponding to permutation $\pi_1$.
In Example~\ref{ex:n4}, four permutations can be excluded because of
optimization~1. Fig.~\ref{data4_1}(b) depicts the matrix after removing
the corresponding four rows. 
\[\begin{array}{ll}
  cover([3,2,1,4])\subseteq cover([3,2,4,1])& \qquad
  cover([3,4,1,2])\subseteq cover([3,4,2,1])\\
  cover([4,2,3,1])\subseteq cover([4,2,1,3])& \qquad
  cover([4,3,2,1])\subseteq cover([4,3,1,2])
  \end{array}
  \]
\noindent{\bf Optimization 2  (graph dominance):}
If graph $G_1$ is covered by a (non-empty) subset of the permutations
that cover graph $G_2$, then we say that $G_1$ dominates $G_2$.
In this case we discard
the column corresponding to permutation $G_2$.
In Fig.~\ref{data4_1}(b), 50 of the 55 graphs are dominated and
discarded. Fig.~\ref{data4_1}(c) depicts the matrix after removing
the corresponding 50 columns and Fig.~\ref{data4_1}(d) depicts the
matrix after applying Optimization~1 one more time.

\smallskip\noindent{\bf Optimization 3 (backbone permutation):}
If a column contains a single one, this corresponds to the case when a
graph is covered by exactly one permutation~$\pi$.  In this case
we say that~$\pi$ is a backbone permutation and must be in any optimal
set-cover. In the terminology of~\cite{roth1969} the row is
``essential''. In this case the row corresponding to~$\pi$ and all
columns corresponding to graphs covered by~$\pi$ are removed.
In Fig.~\ref{data4_1}(d), column 2 contains a single one at row 2,
so row 2 and columns 1, 2 are removed.
Column 5 contains a single one at row 5, so row 5 and columns
4, 5 are removed.
The remaining matrix contains one column with two rows. The column
contains a single one corresponding to a third backbone
permutation. Application of the optimizations renders an empty matrix.

The progression in Fig.~\ref{data4_1}(a--d) illustrates that the
minimal set-cover problem given as Example~\ref{ex:n4} is solved by
repeated application of these three optimizations identifying three
backbone permutations which in this example cover all of the
non-canonical graphs.

\smallskip\noindent{\bf The general case:}
In the general case, after repeated application of the above mentioned
optimizations, we may still need to solve the remaining set-cover
problem.
In our implementation, we  formulate the problem as an \emph{optimization
pseudo-boolean} (OPB) problem and solve it using
RoundingSat~\cite{RoundingSat}.

\smallskip\noindent{\bf Experiment~1:}
Table~\ref{table:results1} summarizes the computation of optimal
lex-constraint symmetry breaks by reduction to the set-cover problem.
For graphs of orders $4$, $5$, and $6$ we compute the matrix
representations of the corresponding set-cover problems and apply the
three optimizations described above. In all three cases the final
matrix is empty and the optimal cover is found.

\begin{wraptable}{r}{7.7cm}
  \vspace{-10mm}
  \caption{Optimal Symmetry Breaks via Set-Cover}\label{table:results1}
  \begin{tabular}{|c|c|c|r|r||r|}
  \hline
    \textsf{order} & \textsf{initial}  & \textsf{cover sizes}
     & \textsf{opt}
    & \textsf{time}~ & \textsf{2016}\\
  \hline
    4     & $23\times 53$     & $24-30$       & 3  & 4.62   & 3  \\
    5     & $119\times 990$   & $448-510$     & 6  & 0.67   & 7  \\
    6     & $719\times 32612$ & $15360-16380$ & 13 & 103.02 & 13 \\
  \hline
  \end{tabular}
  \vspace{-8mm}
\end{wraptable}
The first column details the order of the graph.  The second column
details the size of the initial matrix (rows $\times$ columns). The
third column details the size of the cover sets (number of 1's in each
row).  The forth column (\textsf{opt}) details the size of the minimal
set-cover. The fifth column details the computation time (in
seconds). The sixth column (\textsf(2016)) details the size of the
canonizing set reported in~\cite{Itzhakov2016}.

For graphs of order-7, we  need to construct a matrix of size
$5039\times 2096108$ and manipulate its rows and columns. This task
is beyond the capability of our implementation.
In the following sections we discuss ways to implement the three
optimizations mentioned above before the explicit construction of the
matrix representation of the problem.

\section{A Concise Representation of $cover(\pi)$}

Coming back to Example~\ref{ex:n4}: How do we compute the set of
graphs covered by a permutation $\pi$?
By Def.~\ref{def:cover}, $cover(\pi)$ is the set of solutions of
the constraint $\pi(G)<G$ where $G$ is an unknown graph. We note that
many of these sets tend to contain about half of the non-canonical
graphs.

\begin{example}\label{ex:naive_cover}
  Recall the setting of Example~\ref{example:lex-constraints}.
  The set of graphs covered by the permutation $\pi=[1,2,4,3]$ is the
  set of solutions to the lex-constraint $\pi(G)<G$:
  \begin{equation*}%
   \tuple{x_1,x_3,x_2,x_5,x_4,x_6} <_{lex}
     \tuple{x_1,x_2,x_3,x_4,x_5,x_6}
   \end{equation*}
   which simplifies to
   $\tuple{x_3,x_5} <_{lex} \tuple{x_2,x_4}$ and has the
   following 24 solutions (in decimal representation):
   2, 3, 8, 9, 10, 11, 14, 15, 18, 19, 26, 27, 34, 35, 40, 41, 42,
   43, 46, 47, 50, 51, 58, 59.
\end{example}

We introduce the following to specify that a sequence is lexicographic
smaller than another at position~$i$.  The right-hand side in
Equation~\eqref{eq:smaller-i} is a set of equality constraints which
in our case involve Boolean variables and constants.

\begin{definition}[lexicographic smaller at position $i$]
  Let $\bar a=\tuple{a_1,\ldots, a_m}$ and
  $\bar b=\tuple{b_1, \ldots, b_m}$. Then,
  \begin{equation}
    \label{eq:smaller-i}
    \bar a<_{lex}^i \bar b \Leftrightarrow
    \set{(a_1=b_1),\ldots, (a_{i-1}=b_{i-1}),
      (a_i=0), (b_i=1)}
  \end{equation}
  For a permutation $\pi$, we say that $\pi$ makes a graph $G$ smaller at
  position $i$ if $vec(\pi(G))<_{lex}^i vec(G)$. 
\end{definition}

\begin{observation}
  \begin{equation}\label{eq1}
  \tuple{a_1,\ldots, a_m} <_{lex} \tuple{b_1,\ldots, b_m}
  \Leftrightarrow
  \bigvee\nolimits_{i=1}^{m} \tuple{a_1,\ldots, a_m} <_{lex}^i \tuple{b_1,\ldots, b_m}
\end{equation}
\end{observation}

We now demonstrate that each of the ``$<_{lex}^i$'' constraints in the
disjunction on the right side of Equation~\eqref{eq1} is both
straightforward to solve and  its set of solutions has a
concise representation.

\begin{definition}[patterns]\label{def:patterns}
  Let $\pi$ be a permutation, $G$ be an unknown graph of order-$n$
  with $vec(G)=\tuple{x_1,\ldots,x_m}$ and let $1\leq i\leq m$.
  The pattern, $\pats_i(\pi)$, is the result of applying the most
  general unifier of the equations from the right side in
  Equation~\eqref{eq:smaller-i} to $vec(G)$.
  If the equations have no solution, then $\pats_i(\pi)=\bot$.
  We denote the set of non-$\bot$ patterns corresponding to a
  permutation $\pi$ by $\pats(\pi)$. The patterns corresponding to a
  set of permutations $\Pi$ is denoted $\pats(\Pi)$.
\end{definition}

\begin{example}\label{smart_cover-1}
  Recall $\pi=[1,2,4,3]$ and the constraint
  $\tuple{x_1,x_3,x_2,x_5,x_4,x_6} <_{lex} \tuple{x_1,x_2,x_3,x_4,x_5,x_6}$ from
  Example~\ref{ex:naive_cover}.
  For $i=1,3,5,6$, $\pat_i(\pi)=\bot$. For instance, when $i=1$ the
  equations $\{x_1{=}0,x_1{=}1\}$ have no solution.
  For $i=2$,  applying the most general unifier of
  $\{x_1{=}x_1,x_3{=}0,x_2{=}1\}$ to $\tuple{x_1,x_2,x_3,x_4,x_5,x_6}$ results
  in the pattern $\pat_2(\pi)=\tuple{x_1,1,0,x_4,x_5,x_6}$.
  For $i=4$, applying the most general unifier of
  $\{x_1{=}x_1,x_3{=}x_2,x_2{=}x_3,x_5{=}0,x_4{=}1\}$ to
  $\tuple{x_1,x_2,x_3,x_4,x_5,x_6}$ results in the pattern
  $\pat_4(\pi)=\tuple{x_1,x_2,x_2,1,0,x_6}$.
\end{example}

The solutions of a constraint $\pi(G)<_{lex}G$ are concisely
represented by the corresponding patterns in $\pat(\pi)$.

\begin{example}\label{smart_cover}
  Recall the two patterns detailed in Example~\ref{smart_cover-1} for
  the permutation $\pi=[1,2,4,3]$ (see Table~\ref{table:patterns}). 
  The solutions for the constraint
  $\pi(G)<G$ are obtained as the set of $2^4+2^3$ instances of
  these two patterns.
  These are exactly the 24 solutions specified in
  Example~\ref{ex:naive_cover}.

  \begin{table}[t]
    \caption{Constraints and patterns for
      $\pi(G)<G$ from Examples~\ref{smart_cover-1}
      and~\ref{smart_cover}}\label{table:patterns}
    \centering
    \begin{tabular}{|l|c|c|c|}
      \hline
      \textsf{index} & \textsf{constraint} & \textsf{pattern} & \# \textsf{sols.}\\
      \hline
      $i=2$ & $\{x_1{=}x_1,x_3{=}0,x_2{=}1\}$
                                           & $\tuple{x_1,1,0,x_4,x_5,x_6}$
                                                              & $2^4$\\
      $i=4$ & $\{x_1{=}x_1,x_3{=}x_2,x_2{=}x_3,x_5{=}0,x_4{=}1\}$
                                           & $\tuple{x_1,x_2,x_2,1,0,x_6}$
                                                              & $2^3$\\
      \hline
    \end{tabular}
  \end{table}
 
\end{example}

The set of graphs covered by a permutation $\pi$ may contain an
exponential number of graphs. However, $\pi$ uniquely yields a (small)
set of patterns $\pats(\pi)$ representing the graphs that are covered
by $\pi$.  This means that questions about the sets of graphs covered
by $\pi$ can be addressed on these patterns. 
Phrased as a constraint problem (over the Boolean domain), the set of
 graphs of order-$n$ covered by a permutation $\pi$ is as follows where
 $m=\binom{n}{2}$:
\begin{equation*}\label{eq:pat}
  cover(\pi) = \bigvee\nolimits_{\pat\in\pats(\pi)} \bigwedge\nolimits_{i\in 1..m} x_i = \pat[i]
\end{equation*}
A given graph $G$ with $vec(G)=\tuple{x_1,\ldots,x_m}$ is covered by
$\pi$ if it is an instance of one of the patterns in $\pats(\pi)$.

Observe also that the sets of graphs represented by the different
patterns for a permutation $\pi$ are disjoint: Each pattern denotes
the set of graphs that get smaller ``for the first time'' at a
specified index $i$. It follows that it is easy to compute the number
of graphs covered by a given permutation. For example, by summing the
numbers in the right column of Table~\ref{table:patterns}.

\section{A Symbolic Approach to Set-Cover Optimizations}%
\label{section:permutationDominance}

\begin{figure}[b]
    \centering
    \begin{minipage}{0.3\textwidth}
\begin{tikzpicture}[xscale=.35,yscale=.3
    ]
    \filldraw[thick,fill=red!80!black] (0,1) circle [radius=.1];
    \filldraw[fill=yellow!80!black] (2,1) circle [radius=.1];
    \filldraw[fill=yellow!80!black] (-2,1) circle [radius=.1];
    \filldraw[fill=yellow!80!black] (0,-1.5) circle [radius=.1];
    \draw[very thick] (0,1) ellipse [x radius=3, y radius=1];
    \draw (1,-.5) ellipse [x radius=3, y radius=1,rotate=45];
    \draw (-1,-.5) ellipse [x radius=3, y radius=1,rotate=-45];
    \node at (0,2.75) {$\scriptstyle A$};
    \node at (-2.7, 2.5) {$\scriptstyle B$};
    \node at (+2.7, 2.5) {$\scriptstyle C$};

    \coordinate (C) at (-3.75,-.75);
    \filldraw[fill=yellow!80!black] (C) circle [radius=.1];
    \draw (C) circle [x radius=1,y radius=0.7,rotate=45];
    \draw (C) circle [x radius=2.75,y radius=1,rotate=45];
    \node at (-2.5,-1.75) {$\scriptstyle E$};
    \node at (-4.25,-2) {$\scriptstyle F$};
\end{tikzpicture}

     \end{minipage}
    \hfill
    \begin{minipage}{0.3\textwidth}
\begin{tikzpicture}[xscale=.35,yscale=.3]
    \filldraw[thick,fill=yellow!80!black] (0,1) circle [radius=.1];
    \filldraw[fill=yellow!80!black] (2,1) circle [radius=.1];
    \filldraw[fill=yellow!80!black] (-2,1) circle [radius=.1];
    \filldraw[fill=yellow!80!black] (0,-1.5) circle [radius=.1];
    \draw[very thick] (0,1) ellipse [x radius=3, y radius=1];
    \draw (-1,-.5) ellipse [x radius=3, y radius=1,rotate=-45];
    \node at (0,2.75) {$\scriptstyle A$};
    \node at (-2.7, 2.5) {$\scriptstyle B$};

    \coordinate (C) at (-3.75,-.75);
    \filldraw[fill=yellow!80!black] (C) circle [radius=.1];
    \draw (C) circle [x radius=2.75,y radius=1,rotate=45];
    \node at (-2.5,-1.75) {$\scriptstyle E$};
\end{tikzpicture}

     \end{minipage}
    \hfill
    \begin{minipage}{0.3\textwidth}
\begin{tikzpicture}[xscale=.35,yscale=.3]
    \filldraw[thick,fill=red!80!black] (0,1) circle [radius=.1];
    \filldraw[thick,fill=red!80!black] (2,1) circle [radius=.1];
    \filldraw[fill=yellow!80!black] (-2,1) circle [radius=.1];
    \filldraw[thick,fill=red!80!black] (0,-1.5) circle [radius=.1];
    \draw[very thick] (0,1) ellipse [x radius=3, y radius=1];
    \draw[very thick] (-1,-.5) ellipse [x radius=3, y radius=1,rotate=-45];
    \node at (0,2.75) {$\scriptstyle A$};
    \node at (-2.7, 2.5) {$\scriptstyle B$};

    \coordinate (C) at (-3.75,-.75);
    \filldraw[thick,fill=red!80!black] (C) circle [radius=.1];
    \draw[very thick] (C) circle [x radius=2.75,y radius=1,rotate=45];
    \node at (-2.5,-1.75) {$\scriptstyle E$};
\end{tikzpicture}

     \end{minipage}
    \caption{Backbone and dominance interaction}\label{fig:bbt}
\end{figure}
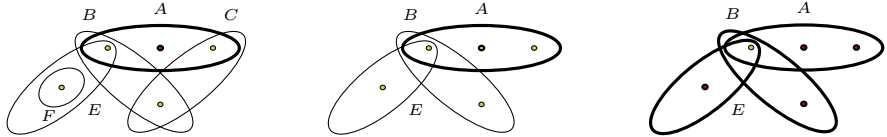

The classic set-cover assumes an explicit representation of
the problem. Here we operate on the elements symbolically by
considering the dominance relation between
permutations and identifying backbones. These two concepts interact,
as exemplified by Fig.~\ref{fig:bbt}, where permutations are depicted
as the sets of graphs that they cover.
Permutation $A$ is a backbone because it covers a graph not covered by
other permutations.
Permutation $E$ is not strictly a backbone because it covers a graph
covered also by $F$, but since $F$ is dominated by $E$, it can be
removed and $E$ becomes a backbone.
Permutations $B$ and $C$ present a more subtle scenario: Neither
appears to be dispensable, \emph{but} if we ignore the graphs already
covered by $A$, they cover the same set of graphs (a single
graph). Hence, it is safe to remove either of them
(non-deterministically). Once $F$ and $C$ are removed, $E$, $B$, and
$A$ are all backbones.

The key observation is that the dominance relation can be considered
\emph{modulo} existing backbones because these must be present in the cover.
Hence, we first detect the set of backbones (see Sec.~\ref{sec:bbones}), then
remove any permutation dominated by another one, modulo the backbones. This is
repeated until a fixed point is found (because new backbones may appear after
the dominance pruning).

\begin{definition}[set of permutations dominates a permutation]
  Let $\Pi$ be a set of permutations and $\pi$ a permutation. If
  $cover(\pi)\subseteq cover(\Pi)$, then we say that $\Pi$ dominates
  $\pi$.
\end{definition}

\begin{definition}[permutation dominates modulo $\beta$]
  Let $\pi_1,\pi_2$ be permutations and $\beta$ a set of (backbone)
  permutations. We say that $\pi_1$ dominates $\pi_2$ modulo $\beta$
  if $cover(\pi_2)\setminus cover(\beta)\subseteq
  cover(\pi_1)\setminus cover(\beta)$.
\end{definition}

Rather than encoding the dominance relation explicitly as a constraint problem,
we develop an encoding via the pattern representation
(Def.~\ref{def:patterns}). This will enable a concise encoding amenable to
incremental solving.

Stating that $\pi_1$ dominates $\pi_2$ modulo $\beta$ is equivalent to
stating that $\beta\cup\{\pi_1\}$ dominates $\pi_2$.
Further, in terms of the pattern representation:
to determine, whether a set of permutations $\Pi$ dominates a
permutation $\pi$, we only need to check that $\pats(\Pi)$
dominates each pattern of $\pats(\pi)$.

To determine, whether a set of patterns~$\Gamma$ dominates a
pattern~$\gamma$, we look for a graph that is \emph{not} covered
by~$\Gamma$ but is covered by~$\gamma$. If such a graph exists,~$\Gamma$
does \emph{not} dominate~$\gamma$.
Hence, for a graph $G$ with $vec(G)=\tuple{x_1,\ldots x_m}$, we wish
to decide the satisfiability of the following sets of constraints
$\textsf{notCovered}(\Gamma)\cup \textsf{covered}(\gamma)$ for each
$\gamma\in\pats(\pi)$.
\noindent
\begin{eqnarray}\label{eq:pat:sub}
  \label{eq:notCovered}    \textsf{notCovered}(\Gamma) =&
     \sset{\hspace{-2mm}\bigvee_{i\in 1..m} x_i \neq \gamma'[i]}{\gamma'\in\Gamma} 
                  &{\texttt{\small\% $G$ is not covered by $\Gamma$}} \\
  \textsf{covered}(\gamma) =& \set{\bigwedge_{i\in 1..m} x_i = \gamma[i]}  
                  &{\texttt{\small\% $G$ is covered by $\gamma$}}
\end{eqnarray}
To encode the problem into SAT, fresh  variables (Tseitin variables~\cite{tseitin68}) are
introduced to represent the constraints of the form $x_i=x_j$, i.e.\
define fresh $e_{ij} \Leftrightarrow (x_i=x_j)$ as 4 clauses
$\{ \lnot e_{ij}\lor\lnot x_i\lor x_j, \lnot e_{ij}\lor x_i\lor\lnot
x_j, e_{ij}\lor x_i\lor x_j, e_{ij}\lor\lnot x_i\lor\lnot x_j \}$.
This encoding is concise and CNF-friendly. The set of graphs covered
by a pattern corresponds to single a conjunction of $e_{ij}$
literals. The set of graphs \emph{not} covered by a pattern
corresponds to a single clause of $\lnot e_{ij}$ literals. The
$e_{ij}$ variables may be reused across multiple patterns and SAT
calls.

\begin{algorithm}[t]
\begin{algorithmic}[1]
  \Procedure{getDominated}{$\Pi$, $\Pi'$}
      \State $R\gets\{\}$; $\Gamma\gets\pats(\Pi)$
      \ForAll{$\pi\in\Pi'$}
      \If{$\bigwedge_{\gamma\in\pats(\pi)}.
        unsat(\textsf{notCovered}(\Gamma)\cup \textsf{covered}(\gamma))$}\label{Line1.4}
            \State $R\gets R\cup\{\pi\}$ 
         \EndIf
      \EndFor
    \State\Return{$R$}
    \EndProcedure
\end{algorithmic}
\caption{Get all permutations in $\Pi'$ dominated by $\Pi$}%
\label{alg:dom}
\end{algorithm}

Alg.~\ref{alg:dom} finds all permutations in a set $\Pi'$
dominated by a set $\Pi$. It relies on the concise
pattern representation to use 
incremental SAT~\cite{minisat} by loading the clauses for
$\textsf{notCovered}(\pats(\Pi))$ into the solver just once.  Each
pattern of a permutation $\pi$ requires a separate SAT call.  However,
if there is a pattern $\gamma\in\pats(\pi)$ not dominated by $\Pi$, then
further SAT calls are unnecessary. Also, satisfiable tests on
Line~\ref{Line1.4} provide
witness graphs, which can be used to prune the \textsf{for all} loop from
other $\pi\in\Pi'$ for which it is also a witness.

The procedure \textsc{getDominated} is readily used in the procedure
\textsc{refine} (Alg.~\ref{alg:refine}), to remove any permutation dominated by
some other permutation in the set modulo the given set of backbones $\beta$.
In Section~\ref{sec:bbones} we will look at algorithms to calculate the set of backbones.

\begin{algorithm}[t]
\begin{algorithmic}[1]
  \Procedure{refine}{$S$,$\beta$}
  \State $S\gets S\setminus\beta$; ~~$T\gets S$ \Comment{permutations to test}
  \While{$T\neq \{\}$}
    \State $\pi\gets\textsf{pick}(T)$; ~~$T\gets T\setminus\{\pi\}$
             \Comment{pick arbitrary permutation from $T$}
    \State $R\gets\textsc{getDominated}(\beta\cup\{\pi\}, S\setminus\{\pi\})$
    \State $S\gets S\setminus R$; ~~$T\gets T\setminus R$ \Comment{remove dominated permutations}
  \EndWhile
  \State\Return{$S$}
  \EndProcedure
\end{algorithmic}
\caption{Refinement of a set $S$ of permutations modulo $\beta$}%
\label{alg:refine}
\end{algorithm}

\section{Backbones for Symmetry Breaking}\label{sec:bbones}

The problem of finding backbones has the nature of one quantifier
alternation, because we are asking whether there \emph{exists} a
permutation $\pi$ that covers some graph~$g$ such that $g$ is not covered
\emph{for all} other permutations $\pi'\neq\pi$. In this section we
explore two approaches to finding backbones: by iterating over all
graphs in Sec.~\ref{sec:bbExplicit}; and by iterating over all
permutations in Sec.~\ref{sec:bbsat}. In the experimentation reported
here we apply a combination of two of these (see
Sec.~\ref{sec:experiments}).

\subsection{Backbone Calculation via Iteration on Graphs}\label{sec:bbExplicit}

The basic algorithm to find permutation backbones is sketched as
Alg.~\ref{findBackbonesAlgorithm}. The set of backbones $\beta$
is initialized to the empty set.  The algorithm iterates over all
graphs $G$ in their $vec(G)$ representation, starting from the empty
graph (all zeroes), and ending with the complete graph (all ones). For
each graph, we test if it is a backbone graph and if so, add a
permutation to $\beta$.
At Line~\ref{line3.5} there is a call to \textsc{backboneStatus}. This
procedure is assumed to return the set $S$ of permutations that cover
$G$ and is refined modulo $\beta$. First, the permutations that cover
$G$ are computed applying an allSAT encoding, and then procedure
\textsf{refine} is applied to remove permutations dominated by others
modulo $\beta$.
The call at Line~\ref{line3.9} increments $vec(G)$ viewing it as a
binary number with least significant bit on the right.
\begin{algorithm}
\begin{algorithmic}[1]
  \Procedure{backbones}{$n$}
  \State initialize $\beta=\emptyset$                    \label{line3.2}
  \State initialize $G = \tuple{0,0,\ldots,0,0}$  \label{line3.3}
  \While{$G\leq \tuple{1,1,\ldots,1,1}$}          \label{line3.4}
  \State $S\gets$\Call{backboneStatus}{$G$,$\beta$} \label{line3.5}
  \If{$S ==\{\pi\}$}                              \label{line3.6} 
  \State $\beta\gets\beta\cup\{\pi\}$
  \EndIf
  \State $G\gets$\Call{increment}{$G$}            \label{line3.9}
  \EndWhile
  \State\Return{$\beta$}
  \EndProcedure
\end{algorithmic}
\caption{find backbones -- the fantasy of iterating over all graphs}%
\label{findBackbonesAlgorithm}
\end{algorithm}

This approach might seem dubious, as there are $2^{\binom{n}{2}}$
graphs to consider along the way. We detail three optimizations that
enable to ``leap'' forward in the iteration such that the algorithm is able
to compute backbones at least for $n\leq 10$.
The first two optimizations are ``exact''. The third is heuristic. It might miss
some backbones but will not wrongly identify a backbone.

\smallskip\noindent{\bf Is $G$ covered by a backbone?}
In the while-loop at Line~\ref{line3.4}, before the expensive call at
Line~\ref{line3.5}, we could test if the current graph $G$ is already
covered by one of the backbone permutations found so far.
In this case we need not check the status of $G$.
This test is efficient given the compact representation $\pat(\beta)$ of
the graphs covered by $\beta$.
Moreover, in this case, the call at line~\ref{line3.9} can be replaced
by a ``leap'' in the iteration. The idea is that if $G$ is covered by
$\beta$, then it is likely that many of the consecutive graphs are
also covered for the same reason.

For example,
let $w=00110$,
consider the order-6 graph $G=w.0^{10}$, and
  assume $G$ is covered by a backbone with  pattern
  $\pat=\tuple{x_1,x_1,x_2,1,0,x_3,\ldots,x_{12}}$.  Observe that any graph
  with the prefix $w$ is also an instance of $\pat$ as the last 10
  elements in $\pat$ consist solely of free variables.  Hence we
  increment $w$ to $w'=00111$ and iteration continues with
  $w'.0^{10}$  thus skipping $2^{10}$ iterations.

\smallskip\noindent{\bf Is $G$ a canonical graph?}
Consider the case where the call at Line~\ref{line3.6} results in
$S=\emptyset$.  This implies that $G$ is canonical as it is not
covered by any permutation.
Again, we can apply a ``leap'' in the iteration at
Line~\ref{line3.9}.
The idea is that if $G$ is canonical and has a suffix of
$k$ zeros, then in may cases, changing a single zero from the suffix
to a one does not make it easier to find a permutation that makes
$G$ smaller.  We test this ``per digit'' and after
$k$ checks we can jump $2^k$ steps in the iteration.

For example,
let
  $w =000001100010101011111$. The order-8 graph $G= w.0^7$ is
  canonical and has a suffix with 7 zeros. With 7 checks, we determine
  that replacing a zero by a one in the suffix does not result in a
  graph that has a cover. Indeed all of the $2^7-1$ graphs succeeding
  $G$ are canonical.  So, we increment $w$ to
  $w'=000001100010101100000$ and iteration continues with $G=w'.0^7$
  thus skipping $2^7$ steps.

\noindent{\bf Does $G$ have a ``huge'' number of covering permutations?}
At Line~\ref{line3.5} in Alg.~\ref{findBackbonesAlgorithm}, in
the call to procedure \textsc{backboneStatus}, we
compute a set $S$ using an allSAT encoding to find all permutations
that cover the graph $G$. As a heuristic, we abort this allSAT
computation if a fixed bound of $B$ permutations are found. We only
apply the \textsc{refine} algorithm if less than $B$ permutations are
found.
Also in this case we can apply a ``leap'' in the iteration at
Line~\ref{line3.9}.
The idea is that when $G$ has a prefix of $k$ zeros, then changing
even a single zero from the suffix to a one, often times, does not make
it easier to  find a permutation that makes $G$ smaller --- so, $G$
will still have a large number of covering permutations.

For example, assume $B=10$ and denote $w=00000000000000000010$. The
graph $G=w.00000000$ has 10 or more covering permutations. After
testing that changing any one of the last 8 zeros to a one results in
a graph that still has at least 10 covering permutations, we increment
$w$ to $w'=00000000000000000011$ and iteration continues with
$G=w'.00000000$ thus skipping $2^8$ steps.

\subsection{Backbone Calculation as a SAT Call}%
\label{sec:bbsat}

Here, we show that it is possible to decide whether a permutation is a
backbone by a SAT call; then we iterate over all permutations to
find all backbones.  The intuition is as follows. All the permutations
cover the whole set of the non-canonical graphs. Removing a backbone
will necessarily diminish the set of covered graphs. In another words,
a backbone $\pi\in S_n$ must \emph{not} be dominated by some set of
permutations $\Pi\subseteq S_n\setminus\{\pi\}$.

Hence, to test whether a given permutation is a
backbone is the test
$\pi\notin\textsc{getDominated}(S_n\setminus\{\pi\}, \{\pi\})$ (see
Alg~\ref{alg:dom}). Such a test needs to be issued for every
permutation, which makes the procedure expensive. This can be
mitigated by quickly eliminating some backbone-candidates by choosing
arbitrarily some set $\Pi\subset S_n$ and marking all
$\textsc{getDominated}(\Pi, S_n\setminus\Pi)$ as non-backbones.%
\footnote{Backbone pruning is also used in simpler, classical SAT
  backbones~\cite{JanotaBB}.}  Some more sophisticated approaches
could be considered, such as in~\cite{Silva13,Belov14}.

This approach  immediately generalizes to an arbitrary set of permutations
$\Pi\subset S_n$, i.e.\ if some permutations were already removed from $S_n$
due to being  dominated, the backbone test becomes
$\pi\notin\textsc{getDominated}(\Pi\setminus\{\pi\}, \{\pi\})$.

We refer to the approach to find backbones described here as
{\bf Algorithm~4}.

\section{Solving the Set-Cover Problems}\label{sec:experiments}

This section puts all of the components together to compute optimal symmetry
breaks. We experimented with the two algorithms described in
Sections~\ref{sec:bbExplicit} (Algorithm~\ref{findBackbonesAlgorithm})
and~\ref{sec:bbsat} (Algorithm~4) to detect backbones. In our trials, the
configuration which works best in practice is to first apply
Alg.~\ref{findBackbonesAlgorithm} and then, starting from this result to apply
the approach from Alg.~4. We alternate the search for backbones with refining
the set of permutations that need to be considered for the set-cover problem.

Table~\ref{table:backbones} provides details of the computation.
The first four columns are about the computation of backbones and
refined permutations. Columns \textsf{bb$_1$} and \textsf{bb$_2$}
detail the number of backbones found after applying
Alg.~\ref{findBackbonesAlgorithm} and then repeatedly applying
Alg.~4 alternating with
Alg.~\ref{alg:refine}.
The final set of permutations that need be considered in the set-cover
problems is detailed in column \textsf{rows}. These are the rows in the
matrix representation.

The next four columns in Table~\ref{table:backbones} are about
creating and solving the matrix representation for the set-cover
problems. The column \textsf{cols} details the number of graphs
covered by the permutations in the rows. The column \textsf{sets}
details the size range of the sets of graphs covered by the
permutations in the rows. The column \textsf{opt} details the size of the
optimal set-cover found, and the column \textsf{2016} details the number
of permutations in the canonizing sets reported
in~\cite{Itzhakov2016}.

The next three columns are about times (rounded to seconds, minutes,
or hours): \textsf{time$_1$} to compute backbones \textsf{bb$_1$};
\textsf{time$_2$} to compute \textsf{bb$_2$} and \textsf{rows}; and
\textsf{time$_3$} to generate the matrix representation,
apply Optimizations~1--3, and solve the set-cover problem.
Interestingly, for $n<10$ we never needed to solve the general
set-cover because everything is solved by repeated applications of the
described optimizations.
For $n=10$, the optimizations reduce the initial matrix to one with
199 rows and 197 columns.  We formulate the remaining set-cover
problem as an \emph{optimization pseudo-boolean} (OPB) problem and
solve it by \textsf{RoundingSat}~\cite{RoundingSat}.
This OPB problem %
is solved in 0.01s.

The final column (\textsf{enc.~size}) shows the number of clauses to
encode the symmetry break; this is done by negating all the patterns
(see Equation~\eqref{eq:notCovered}) in the optimal cover---the
pattern encoding is 3 orders of magnitude smaller than the direct
lex-leader encoding~(see Def.~\ref{def:canonizing}).

All of the complete symmetry breaks found in this paper were verified,
using GANAK~\cite{SharmaEtAl19a}, to have a number of solutions
corresponding to the number of non-isomorphic graphs (sequence A000088
of the OEIS~\cite{OEIS}).

\begin{table}[t]
  \caption{Constraints and patterns for
      $\pi(G)<G$ from Examples~\ref{smart_cover-1}
      and~\ref{smart_cover}}\label{table:backbones}
  \centering
\begin{tabular}{|c||r|r|r||r|l|r|r||r|r|r||r|}
  \hline
  \textsf{order}&\textsf{~bb$_1$}&\textsf{bb$_2$~~}&\textsf{rows}&
  \textsf{cols~~}&\textsf{~sets}&\textsf{opt}&2016&
  \textsf{time$_1$}&\textsf{time$_2$}&\textsf{time$_3$} & \textsf{enc.~size}\\
  \hline
  6  & 9  &  13 &   0&   0  & 0-0  &  13&  13&   89s&    0s&  0s & 463\\
  7  & 18 &  25 &  44& 148  & 3-72 &  35&  37&  123s&    1s&  0s & 956\\
  8  & 30 & 112 &  19&  33  & 3-10 & 121& 135&  929s&   55s&  0s & 1,925\\
  9  & 90 & 709 & 207& 344  & 2-42 & 765& 842&  289m&   75m& 58s & 5,190\\
  10 & 131&6,920 & 694& 3,481 & 2-550&7,181&7,853&  613m&  137h& 68m & 31,193\\
  \hline
\end{tabular}
\end{table}

\section{Backbones as a Partial Symmetry Break}
\begin{wraptable}{r}{4.4cm}
  \vspace{-1.1cm}
  \caption{ Partial symmetry breaks: transpositions vs.\ backbones } \label{table:partial}
  \begin{tabular}{|c|c|c|c|c|c|}
    \hline
    \textsf{order} & \textsf{trns} & $\rho(\textsf{trns})$
    & \textsf{bb}$_1$ & $\rho(\textsf{bb}_1)$  \\
    \hline
    6  & 15 & 1.76  & 9  & 1.19 \\
    7  & 21 & 3.02  & 18 & 1.36 \\
    8  & 28 & 5.39  & 30 & 1.87 \\
    9  & 36 & 9.42  & 90 & 1.99 \\
    10 & 45 & 15.34 &131 & 2.99 \\
    \hline
  \end{tabular}
   \vspace{-0.7cm}
\end{wraptable}
Codish~\etal\cite{DBLP:conf/ijcai/CodishMPS13,Codish2019} introduce a
polynomial sized partial symmetry breaking constraint for graphs
defined in terms of transpositions (permutations that swap two
vertices).
In fact, any set of permutations $\Pi$ can be viewed as a
partial symmetry break obtained by replacing $S_n$ with $\Pi$ in
Equation~\eqref{lexleader}.
Heule defines the notion of \emph{redundancy ratio} to measure the
precision of symmetry breaks on graphs~\cite{Heule2019}. 
In our
terminology, the redundancy ratio for a set of permutations $\Pi$,
which we denote $\rho(\Pi)$, is the ratio between the number of graphs
that are not covered by $\Pi$ and the number of isomorphism
classes. If $\Pi$ is canonizing, then $\rho(\Pi)$ is 1.
Table~\ref{table:partial} details the numbers, \textsf{trns} and
\textsf{bb}$_1$, of transpositions and backbones found using
Alg.~\ref{findBackbonesAlgorithm}, together with the redundancy
ratios $\rho(\textsf{trns})$ and $\rho(\textsf{bb}_1)$.
One can observe that the symmetry break using backbones involves a
relatively small set of permutations, and provides a partial
symmetry break which is  more precise.

\section{Conclusion}

This paper formalizes symmetry breaking as a set-covering problem and
this is the main contribution of the paper. As demonstrated by the paper,
this formalization opens up a range of new
opportunities for complete and partial symmetry breaking deriving from
decades of studies on both precise and approximate techniques for this
problem.
We focus primarily on precise solutions to provide complete symmetry
breaks for graphs which are optimal in the number of lex-constraints.
We achieve this for all cases in which small complete
symmetry breaks based on lex-constraints have been computed
in~\cite{Itzhakov2016}. Namely, for graphs of order $n\leq 10$.
Interestingly (see Table~\ref{table:backbones}), the symmetry breaks
computed in~\cite{Itzhakov2016} are less than 10\% larger than the
optimal ones.

An important ingredient is the notion of \emph{patterns} to provide a
concise representation of the (possibly exponential) set of graphs
covered by a permutation.
Another important ingredient is in the notion of a \emph{backbone},
which is a permutation that is the only one that covers some graph
that is not already covered by other permutations.
We apply two types of optimizations: \emph{before} the construction of
the matrix representation of the set-cover problem; these are symbolic
and rely on SAT encodings, and \emph{after} the construction.  Both
types identify permutations and graphs that can be ignored in the
search for an optimal cover, and both types identify backbone
permutations.

It is interesting to note that for graphs of order $n<10$, our
construction of the optimal symmetry breaks never needed to solve a
general set-cover problem.  Everything is solved by repeated
application of the described optimizations (before and after the
construction of the matrix representation).
For the case $n=10$, the original set-cover problem of size
$(10!\times 2^{45})$ is finally reduced to a set-cover problem of size
$(199\times 197)$.  In all cases, the repeated identification of
backbones can be seen as ``driving'' the construction.

In this paper we consider set-covers defined in terms of the sets of
graphs covered by permutations. It is interesting to
investigate set-covers defined in terms of patterns (which are similar
to the implications defined in~\cite{DBLP:conf/flops/CodishEGIS18})
and in terms of ``isolaters'' as applied in~\cite{Heule2019}.

Encouraged by the results in Table~\ref{table:partial}, where a small
set of backbone permutations is shown to provide a strong partial
symmetry break, it is interesting to further investigate the application of
set-cover techniques to construct small and precise partial symmetry
breaks.

\begin{credits}
\subsubsection{\ackname}
The results were supported by the Ministry of Education, Youth and Sports
within the dedicated program ERC~CZ under the project \emph{POSTMAN} no.~LL1902
and co-funded by the European Union under the project \emph{ROBOPROX}
(reg.~no.~CZ.02.01.01/00/22\_008/0004590).
This article is part of the \emph{RICAIP} project that has received funding
from the European Union's Horizon~2020 research and innovation programme under
grant agreement No~857306.

\subsubsection{\discintname}
The authors have no competing interests to declare that are
relevant to the content of this article.
\end{credits}

\bibliographystyle{splncs04}

\end{document}